\documentstyle[12pt,epsfig]{article} 
\begin{document}
\title{Decoherence--induced violations of Einstein equivalence principle}
\author{ A. Camacho
\thanks{email: acamacho@nuclear.inin.mx}
\thanks{This essay received an ``honorable mention'' in the 2001 Essay Competition of the Gravity Research Foundation--Ed.} \\
Department of Physics, \\
Instituto Nacional de Investigaciones Nucleares,\\
Apartado Postal 18--1027, M\'exico, D. F., M\'exico.}

\date{}
\maketitle

\begin{abstract}
In this essay it will be shown that Decoherence Model and Einstein Equi\-va\-lence Principle are 
conceptually incompatible.
In other words, assuming only the validity of the Weak Equivalence Principle the present work concludes that we face two possibi\-lities: (i) if Decoherence Model provides a co\-rrect description of nature at quantum level,
then there are systems which violate Local Position Invarian\-ce, or, (ii) if all the postulates behind Einstein Equi\-va\-lence Principle
are valid, even on quantum realm, then Decoherence Model breaks down in curved spacetimes.
Finally, the present results are confronted against Schiff's conjecture.
\end{abstract}
\newpage
\section{Introduction}
\bigskip

Nowadays the number of experiments that confirm the predictions of General Relati\-vi\-ty (GR) seems to be 
overwhelming [1, 2]. From the fact that Einstein Equivalence Principle (EEP) is the cornerstone of any metric 
theory of gravity [1], we may conclude that many of the aforementioned experiments embody also an indirect test of EEP. 
At this point it is noteworthy to comment that there are also experiments that confront, against measurement outputs, some of the postulates of EEP. 
For instance, experimental tests of the Weak Equivalence Principle (WEP) [3], or of Local Position Invariance (LPI) [4, 5].

This huge amount of experimental data could lead us to take for granted the validity of EEP, not only on classical level, but also 
on the quantum one. Neverwithstanding, recently some papers have appeared in which
even the simplest kinematical ideas show conceptual difficulties when they
are extrapolated to the quantum realm. For example, there is no unique definition of
the probability distribution for the time of flight in a homogeneous gravi\-tational field of a quantum system
without classi\-cal analogue [6].
Even more, we may already find works claiming the violation of EEP,
either in connection with systems that have no classical analogue [7], or with situations in which a quantum measuring process 
plays an important role [8].

On the other hand, one of the more long--standing conundrums in Quantum Theo\-ry (QT) comprises the so called quantum measurement issue, which has several intriguing aspects, one of them is the question of how a classical world emerges from QT. There are several models which claim to give an answer to this conceptual difficulty [9], and that hold incompatible points of view. 
Nevertheless, we may find some approaches which are equivalent [10], 
and which involve, in one way or in another, the so called decoherence process [11].

In the present essay the relation between EEP and Decoherence Model (DM)
will be analyzed, namely, we will consider the following question: {\bf assuming only the validity of WEP, is DM conceptually consistent with EEP?}

In the search for an answer we proceed as follows. The relation between DM and EEP will be explored considering the quantum mechanical superposition of diffe\-rent mass eigenstates, for example, two neutrinos related with two different 
lepton genera\-tions, a so called flavor--oscillation clock [12]. Employing the master equation [10], a measuring process for 
the energy of this system will be introduced, and the corresponding differential equations for the elements of the density matrix will be solved. 
Afterwards, taking for granted only the validity of WEP, it will be proved that if we accept DM, then EEP is, unavoidably, violated. Indeed, either the off--diagonal elements of the corresponding density matrix do not satisfy LPI, or the measuring device must contain always information concerning the gravitational field. This last possibility could be reworded asserting that any experiment performed in the respective local Lorentz frame is always influenced by the gravitational background, 
a possibility already pointed out in the context of the analysis of the behavior of some of the symmetries behind a continuous quantum measurement [13]. 

Our main conclusion will be that decoherence could be a source of violations of EEP, in particular of LPI, and that these violations could be, in some cases, detected, in principle, measuring the time evolution of the off--diagonal elements of the density matrix. 

In other words, the answer to our initial question reads: if only the validity of the Weak Equivalence Principle is assumed, then we face two alternatives; (i) if DM provides a correct description of nature at quantum level, then there are systems which violate LPI, (ii) if all the postulates behind EEP are valid, even on quantum realm, then DM breaks down in curved spacetimes.
\bigskip
\bigskip

\section{Decoherence and flavor--oscillation clocks}
\bigskip
\bigskip

Let us consider the quantum mechanical superposition of different mass eigenstates, 
for example, two neutrinos related with two different lepton generations

{\setlength\arraycolsep{2pt}\begin{eqnarray}
|F_1; t = t_0>~= \cos(\theta)|m_1> + \sin(\theta)|m_2>,
\end{eqnarray}}

{\setlength\arraycolsep{2pt}\begin{eqnarray}
|F_2; t = t_0>~= -\sin(\theta)|m_1> + \cos(\theta)|m_2>.
\end{eqnarray}}

Clearly these two kets are orthogonal to each other.

In our case the background geometry is described by the following line element

{\setlength\arraycolsep{2pt}\begin{eqnarray}
ds^2 = -(1 - {2GM\over rc^2} - 2\vert\phi\vert)dt^2 + (1 + {2GM\over rc^2} + 2\vert\phi\vert)(dx^2 + dy^2 + dz^2),
\end{eqnarray}}

\noindent where $\phi$ is a non--vanishing constant term and $r$ is the distance to the center of a spherical body with 
mass $M$. We also have that $0 <|\phi| <<1$, this is a condition that has to be fulfilled in
order to have a weak field approximation [14].

This constant contribution, $\vert\phi\vert$, to the gravitational potential
could have physical meaning, for instance, it could stem from the
gra\-vi\-tational potential of the local cluster of galaxies, the so called
Great Attractor [15]. Concerning its properties in the solar system, this potential is constant
to about 1 part in $10^{11}$ [7].

At this point the consistency of (3) with Einstein's equations must be addressed. 
If we calculate, up to linear order in ${GM\over rc^2}$ and $\vert\phi\vert$,
Riemann, Ricci, and Einstein tensors, stemming from (3), and compare them with the
corresponding variables when the term $\vert\phi\vert$ is absent, then we find that these
tensors are the same. But the case when $\vert\phi\vert$ is absent comprises a solution to the linearized Einstein equations [14],
hence expression (3) is also a solution of these equations. 
In other words, the metric that our line element defines is consistent with the linearized Einstein equations.

We will also assume that our two quantum systems have vanishing small three--momentum, 
i.e., they are at rest with respect to the coordinate system defined by the line element given in (3).
Under these conditions the Hamiltonian reads [16] 

{\setlength\arraycolsep{2pt}\begin{eqnarray}
H = \left[1 - {GM\over rc^2} - \vert\phi\vert~\right]c^2\tilde{M}.
\end{eqnarray}}

Here $\tilde{M}$ is the mass operator, i.e., $\tilde{M}|m_1> = m_1|m_1>$, and $\tilde{M}|m_2> = m_2|m_2>$.

Let us now suppose that we measure the energy of the state whose initial ket is given by (1). This measuring process 
may be described employing the master equation formalism [10] for the density operator. In our case this equation is

{\setlength\arraycolsep{2pt}\begin{eqnarray}
\dot{\rho} = -{i\over \hbar}[H, \rho] - {\kappa\over 2}[H, [H, \rho]].
\end{eqnarray}}

The master equation is related to the so--called nonselective approach, in which no particular history of measurement is selected. The first term in the right hand side of expression (5) results from the dynamics of the free situation, while the second one provides the decay of the off--diagonal matrix elements. Here $\kappa$ represents the coupling of the energy meter to the system, and, in principle, it may not only depend upon time but also be an operator (having the same eigenvalues as the Hamiltonian). In the context of the restricted path integral formalism (a model that in the case of nonselective measurements is equivalent to the master equation) we may interpret $\kappa$ as the width of the quantum corridor. In other words, $\kappa$ is connected with the measurement resolution of the experimental device, i.e., it characterizes the properties of the measuring apparatus [17].

The initial value of this density operator is 

{\setlength\arraycolsep{2pt}\begin{eqnarray}
\rho(0) ~= \left(\cos(\theta)|m_1> + \sin(\theta)|m_2>\right)\left(\cos(\theta)<m_1| + \sin(\theta)<m_2|\right).
\end{eqnarray}}

Denoting $\rho_{ij} = <m_i|\rho|m_j>$, with $ i, j = 1, 2$, the corresponding differential equations are

{\setlength\arraycolsep{2pt}\begin{eqnarray}
\dot{\rho}_{11}(t) = 0,
\end{eqnarray}}

{\setlength\arraycolsep{2pt}\begin{eqnarray}
\dot{\rho}_{22}(t) = 0, 
\end{eqnarray}}
 
{\setlength\arraycolsep{2pt}\begin{eqnarray}
\dot{\rho}_{12}(t) = \left(m_1 - m_2\right)U\left[- {i\over\hbar} - {\kappa(t)\over 2}(m_1 - m_2)U\right]{\rho}_{12}(t),
\end{eqnarray}}

{\setlength\arraycolsep{2pt}\begin{eqnarray}
\dot{\rho}_{21}(t) = \left(m_2 - m_1\right)U\left[- {i\over\hbar} - {\kappa(t)\over 2}(m_2 - m_1)U\right]{\rho}_{21}(t),
\end{eqnarray}

{\setlength\arraycolsep{2pt}\begin{eqnarray}
U = \left[1 - {GM\over rc^2} - \vert\phi\vert~\right]c^2.
\end{eqnarray}

As already mentioned, $\kappa$ represents the coupling of the energy meter to the system, and, in principle, it could also depend upon time.

It is readily seen that 

{\setlength\arraycolsep{2pt}\begin{eqnarray}
|\rho_{11}(t)|^2 = |\rho_{11}(0)|^2,
\end{eqnarray}

{\setlength\arraycolsep{2pt}\begin{eqnarray}
|\rho_{22}(t)|^2 = |\rho_{22}(0)|^2,
\end{eqnarray}

{\setlength\arraycolsep{2pt}\begin{eqnarray}
|\rho_{12}(t)|^2 = \exp\left\{-\left[(m_1 - m_2)U\right]^2\int_{0}^{t}\kappa(\tau)d\tau\right\}|\rho_{12}(0)|^2,
\end{eqnarray}

{\setlength\arraycolsep{2pt}\begin{eqnarray}
|\rho_{21}(t)|^2 = \exp\left\{-\left[(m_2 - m_1)U\right]^2\int_{0}^{t}\kappa(\tau)d\tau\right\}|\rho_{21}(0)|^2.
\end{eqnarray}

In a Minkowskian spacetime we have that 

{\setlength\arraycolsep{2pt}\begin{eqnarray}
U = c^2.
\end{eqnarray}

Therefore (14) becomes (henceforth we consider only $|\rho_{12}(t)|$, but the analysis of $|\rho_{21}(t)|$ leads to the same conclusions)

{\setlength\arraycolsep{2pt}\begin{eqnarray}
|\rho^{(F)}_{12}(t)|^2 = \exp\left\{-\left[(m_1 - m_2)c^2\right]^2\int_{0}^{t}\kappa(\tau)d\tau\right\}|\rho_{12}(0)|^2.
\end{eqnarray}

In order to analyze the predictions of GR we resort now to WEP, i.e., we consider 
now a locally inertial frame. The condition of vanishing three--momentum means that
we must choose that locally inertial frame in which the oscillation clock is momentarily
at rest.
As has already been pointed out [7], WEP allows
us to annul the gradients of the gravitational potential, but it can not discard 
its constant parts. This remark means that if we resort to WEP in expression (14), 
then we may annul ${GM\over c^2r}$ but $\vert\phi\vert$ remains. 
At this point we must underline that we are not allowed to employ the transformation rule that expresses the coordinates of the freely falling frame in terms of those appearing in (3), for instance, according to this rule the time coordinate in a freely falling frame, $t^{(f)}$, is given by $t^{(f)} = t/\sqrt{1 - {2GM\over rc^2} - 2\vert\phi\vert}$. Neverwithstanding, implicitly, behind this expression lies the fact that the ticking rate of the clocks, located in our freely falling reference frame, coincides with the predictions of special relativity, i.e.,  we impose the validity of special relativity, which means that we have also introduced EEP, but we are allowed only to assume WEP, which states that we may get rid only of gradients of the gravitational potential.  

Hence if only WEP is accepted, then $U$, in our freely falling reference frame, reads

{\setlength\arraycolsep{2pt}\begin{eqnarray}
U = \left[1 - \vert\phi\vert~\right]c^2.
\end{eqnarray}

Therefore, 

{\setlength\arraycolsep{2pt}\begin{eqnarray}
|\rho^{(L)}_{12}(t)|^2 = \exp\left\{-\left[(m_1 - m_2)\left[1 - \vert\phi\vert~\right]c^2\right]^2\int_{0}^{t}\tilde{\kappa}(\tau)d\tau\right\}|\rho_{12}(0)|^2.
\end{eqnarray}

Here we have assumed that the initial values in the Minkowskian case, and in a locally freely falling coordinate system are the same. The coupling parameter in
the case of a locally Lorentz frame has been written as $\tilde{\kappa}$. The reason for this lies on the fact that we do not assume from the begining that EEP is valid, 
which would be the case if in (19) instead of $\tilde{\kappa}$ appears $\kappa$, i.e., we only accept WEP.
\bigskip

Let us now address the relation of EEP with the previous results. At this point we face two possibilities:
\bigskip
\bigskip

(1) The values of the off-diagonal elements coincide, i.e., $|\rho^{(F)}_{12}(t)|^2 = |\rho^{(L)}_{12}(t)|^2$. 
But in this case, a necessary condition for the fulfillment of this equaliy is

{\setlength\arraycolsep{2pt}\begin{eqnarray}
\int_{0}^{t}\tilde{\kappa}(\tau)d\tau = {1\over 1 - \vert\phi\vert}\int_{0}^{t}\kappa(\tau)d\tau.
\end{eqnarray}

This last expression implies that the coupling parameter between experimental device and measured system contains information 
about the gravitational potential $\phi$. Put it another, it does not coincide with the value of the corresponding case in a Minkowskian spacetime, 
and in consequence, EEP is violated. More precisely, it is readily seen that it
depends upon the value of $\phi$, and in consequence LPI is violated.
\bigskip

(2) The coupling parameter does not depend upon the gravitational potential $\phi$

{\setlength\arraycolsep{2pt}\begin{eqnarray}
\int_{0}^{t}\tilde{\kappa}(\tau)d\tau = \int_{0}^{t}\kappa(\tau)d\tau.
\end{eqnarray}

Introducing this in (19) we conclude that 

{\setlength\arraycolsep{2pt}\begin{eqnarray}
|\rho^{(F)}_{12}(t)|^2 \not= |\rho^{(L)}_{12}(t)|^2.
\end{eqnarray}

Once again we end up with a violation of EEP, and once again this fact appears in connection with LPI, i.e., expression (22) depends upon the value of $\phi$.

Let us now address the order of magnitude of the departure from the general--relativistic predictions for a very particular case. Suppose now that 
$\phi$ stems from the Great Attractor [15], then, $\phi \sim 10^{-5}$ and that $|m_1 - m_2| \sim 10^{-1}$ e--V [7].

Under these conditions, if the measuring device does not depend upon $\phi$, then we may detect the violation of EEP monitoring the off--diagonal elements of the density matrix

{\setlength\arraycolsep{2pt}\begin{eqnarray}
|\rho^{(F)}_{12}(t)|^2  = \left(|\rho^{(L)}_{12}(t)|^2\right)^{(1 -  \vert\phi\vert)^2} \sim \left(|\rho^{(L)}_{12}(t)|^2\right)^{(0.9994)}. 
\end{eqnarray}}

\bigskip
\bigskip

\section{Conclusions}
\bigskip
\bigskip

Considering the density operator associated with the superposition of two different mass eigenstates of the neutrino, 
we have analyzed the effects of a measuring process upon this system. This has been done employing the master equation, 
and then solving the corresponding differential equations. Assuming only the validity of WEP, it has been found that a violation of EEP will always emerge, namely, either the off--diagonal elements of the density matrix do not satisfy LPI, or any local experiment carried out in the respective local Lorentz frame is influenced by the gravitational background, 
and in consequence, once again, LPI breaks down. In other words, though WEP is valid, the simultaneous use of DM and EEP leads to contradictions.
The order of magnitude of the departure from the general relativistic predictions, in a possible terrestrial experiment, was also addressed. 

An important point is that though we have considered the measuring process of this superposition of mass eigenstates, the master equation may also 
represent the interaction of our system with its environment [17]. In other words, in this new interpretation $\kappa$ and $\tilde{\kappa}$ would represent not the 
interaction with an energy meter, but the coupling of the energy of this superposition with its environment.
This last fact means that in more realistic schemes (than those hitherto considered [7, 8, 12]) EEP violations do not disappear, i.e.,
DM shows us that the environment, in which our system is embeded, does not restore the validity of EEP.  This last remark also shows a new element that the present analysis contains, and which is absent in previous results [7, 12], namely, even though the possible violation of EEP has already been considered in the context of flavor--oscillation clocks, the consequences on this scenario of the corresponding  environment have not yet been analyzed.

We have taken for granted the validity of WEP (freely falling coordinate systems coincide with locally Lorentz frames) and, nevertheless, violations of EEP have been found. 
This allows us to assert that Schiff's conjecture ({\it any complete, self--consistent theory of gravity that embodies WEP necessarily embodies EEP}) 
[1, 18] could be, at least on the quantum realm, false. 

It is interesting to confront the present work with some of the arguments in favor 
of Schiff's conjecture, here we follow section 2.5 of [1]. We have that a possible violation of LPI (which implies a violation of WEP) stems from an 
anomalous passive gravitational mass term, which 
has been deduced considering a quantum system, the one goes from a state, that is an energy eigenstate, to another one, which also is an energy eigenstate.  
In other words, the only source of LPI violations is a parameter that emerges from quantum--mechanical arguments which from the very begining 
discard the possible consequences of the superposition of energy eigenstates.
Clearly the present essay contemplates a possibility that lies beyond the vali\-di\-ty
region of the assumptions that support Schiff's conjecture.

Summing up, we may conclude (assuming only the validity of the Weak Equi\-valence Principle) that there are two alternatives; (i) if Decohe\-ren\-ce Model provides a correct description of nature at quantum level, then there are systems which violate
Local Position Invarian\-ce, (ii) if all the postulates behind Einstein Equivalence Principle
are valid, even on quantum realm, then Decoherence Model breaks down in curved spacetimes.
Hence the answer to our original question (see first section) reads: {\bf no,
Decoherence Model and EEP can not be simultaneous\-ly valid}.

\bigskip
\bigskip

\Large{\bf Acknowledgments.}\normalsize
\bigskip

The author would like to thank A. A. Cuevas--Sosa for his help. This work was partially supported by CONACYT (M\'exico) Grant No. I35612--E.

\end{document}